\documentclass[aps,prb,twocolumn,superscriptaddress,showpacs]{revtex4-1}

\usepackage{hyperref}
\usepackage[dvips]{graphicx}
\usepackage{amssymb,amsmath,amsfonts}
\usepackage[T1]{fontenc}
\newcommand{\rethree}{$(\sqrt{7}\times\sqrt{7})$R$19.1^{\circ}$}

\newcommand{\ke}{($\vec{k}, E$)}
\newcommand{\mpar}{$s_\parallel$}
\newcommand{\mper}{$s_\perp$}
\newcommand{\auad}{Au$_{\mathrm{ad}}$}
\newcommand{\ptad}{Pt$_{\mathrm{ad}}$}
\newcommand{\auin}{Au$_{\mathrm{in}}$}
\newcommand{\ptin}{Pt$_{\mathrm{in}}$}

\begin{document}


\title{Spin-orbit proximity effect in graphene on metallic substrates: decoration vs intercalation with metal adatoms}


\author{Jagoda S\l awi\'{n}ska}
\affiliation{Present address: Department of Physics, University of North Texas, Denton, TX 76203, USA}
\affiliation{Instituto de Ciencia de Materiales de Madrid, ICMM-CSIC, Cantoblanco, 28049 Madrid, Spain}
\affiliation{Department of Solid State Physics, University of \L \'{o}d\'{z}, Pomorska 149/153, 90236 \L \'{o}d\'{z}, Poland}
\author{Jorge I. Cerd\'{a}}
\affiliation{Instituto de Ciencia de Materiales de Madrid, ICMM-CSIC, Cantoblanco, 28049 Madrid, Spain}

\vspace{.15in}

\date{\today}

\begin{abstract}
The so-called spin-orbit proximity effect experimentally realized in graphene (G) on
several different heavy metal surfaces opens a new perspective to engineer the spin-orbit coupling (SOC) for new generation spintronics devices. Here, via large-scale density functional theory (DFT) calculations performed for two distinct graphene/metal models, G/Pt(111) and G/Au/Ni(111), we show that the spin-orbit splitting of the Dirac cones (DCs) in these stuctures might be enhanced by either adsorption of adatoms on top of graphene (decoration) or between the graphene and the metal (intercalation). While the decoration by inducing strong graphene-adatom interaction suppresses the linearity of the G's $\pi$ bands, the intercalated structures reveal a weaker adatom-mediated graphene/substrate hybridization which preserves well-defined although broadened DCs. Remarkably, the intercalated G/Pt(111) structure exhibits splittings considerably larger than the defect-free case.
\end{abstract}

\maketitle
\section{Introduction}
Tuning of spin-orbit coupling (SOC) in graphene \cite{spintronics} is one of the fundamental steps to engineer graphene-based spintronics devices. One promising route to achieve this goal is the so-called spin-orbit proximity effect, recently extensively studied from both theoretical and experimental side.\cite{proximity, fluorine, weeks, tmds, njp, smallcells, copper, carbon, soc_prb, voloshina,krivenkov} This mechanism of inducing SOC extrinsically relies on the proximity between graphene (G) and a metal;
the SOC of the heavy atoms might be \textit{transferred} to the G when both materials are brought
sufficiently close to each other. Experimental realizations of spin-orbit proximity have revealed
several important phenomena, such as spin Hall effect at room temperature shown by Avsar
\textit{et al.}~\cite{proximity} or even a more intriguing electron confinement associated to
multiple topologically non-trivial gaps observed by Calleja \textit{et al.} in graphene
on Ir intercalated by Pb nanoislands (Pb/Ir).\cite{calleja}

Recently, we have reported that the mechanism of inducing SOC in G when adsorbed on heavy metal
surfaces is far more complex than it had been predicted before.\cite{soc_prb} DFT calculations of
graphene on Pt(111) and on Au/Ni(111) showed that the induced spin texture is a result of
spin-dependent hybridization between the Dirac cones (DCs) and the surface $d$-bands of the metal.
The spin vector of graphene is determined by that of the substrate bands, and undertakes
rotations wherever hybridization with any of the spin-orbit splitted metal bands occur.
Consequently, the reported non-trivial spin textures, although intriguing from the
fundamental point of view, seem difficult to control in any practical device.
Furthermore, although hybridizations locally open mini-gaps around which the SOC-derived spin
splitting may reach giant values above 100~meV, in the quasi-linear regions, where
the G transport properties are most relevant, the splittings are typically of the order of
just 10~meV.\cite{soc_prb, voloshina}

The main purpose of this study is to theoretically explore alternative routes to increase the
SOC derived splittings in the G by incorporating single metal adatoms at the graphene/metal
interface.  We consider two types of adsorption which should lead to two very different
interaction scenarios:
(i) decoration defined as the adsorption of the adatom on top of graphene and,
(ii) intercalation of the adatom between the G and the metallic surface.
The first case should induce changes mainly in the G's properties already perturbed by the metal
surface, while the latter might
significantly alter the graphene-substrate proximity, as graphene will now interact with the
metal mainly via the intercalated adatom. Importantly, both decoration and intercalation can be realized experimentally\cite{decoration1, decoration2, decoration3, decoration4, intercalation1, intercalation2, intercalation3, calleja, misha, jorge, 2dmaterials, krivenkov} and are known to provide several interesting options for engineering of graphene's properties, in addition to any possible enhancement of SOC.\cite{krajl} Here, we will focus on two previously studied models, G/Pt(111) and G/Au/Ni(111) which present markedly different electronic and magnetic properties, and consider the adsorption of one species for each system, namely, a Pt adatom for G/Pt(111) and
an Au adatom for G/Au/Ni(111).

The paper is organized as follows. In Sec II we provide a brief description of DFT calculations. Section III reports the electronic properties and spin textures of the G/Pt(111) calculated defect-free case and both types of adsorption. In Sec. IV we present a similiar analysis for G/Au/Ni(111) structures. The conclusions are summarized in Sec. V.

\section{Methods}
Our large-scale DFT calculations have been performed with the {\sc SIESTA} code~\cite{siesta} as implemented within the {\sc GREEN} package.\cite{green,loit} The exchange-correlation (XC) potential has been treated using the generalized gradient approximation (GGA) in the Perdew, Burke, and Ernzerhof formalism.\cite{pbe} Dispersion forces were included via the
semi-empirical scheme of Ortmann and Bechstedt.\cite{vdw_ortmann} Spin-orbit coupling has been self-consistently taken into account as implemented in Ref. \onlinecite{soc}. Core electrons have been simulated employing norm-conserving pseudopotentials of the Troulliers-Martin type, including core corrections for the metal atoms.
The atomic orbital (AO) basis set based on double-zeta polarized strictly localized numerical
orbitals has been generated employing a confinement energy of 100~meV.
Real space three-center integrals have been computed over 3D-grids with a resolution equivalent to
500~Rydbergs mesh cut-off, while the Brillouin zone integrations have been performed over
$k$-supercells of around ($18\times18$) with respect to the G-($1\times1$) unit cell. The
temperature $kT$ in the Fermi-Dirac distribution has been set to 10 meV in all cases.

\begin{figure*}
\includegraphics[width=\textwidth]{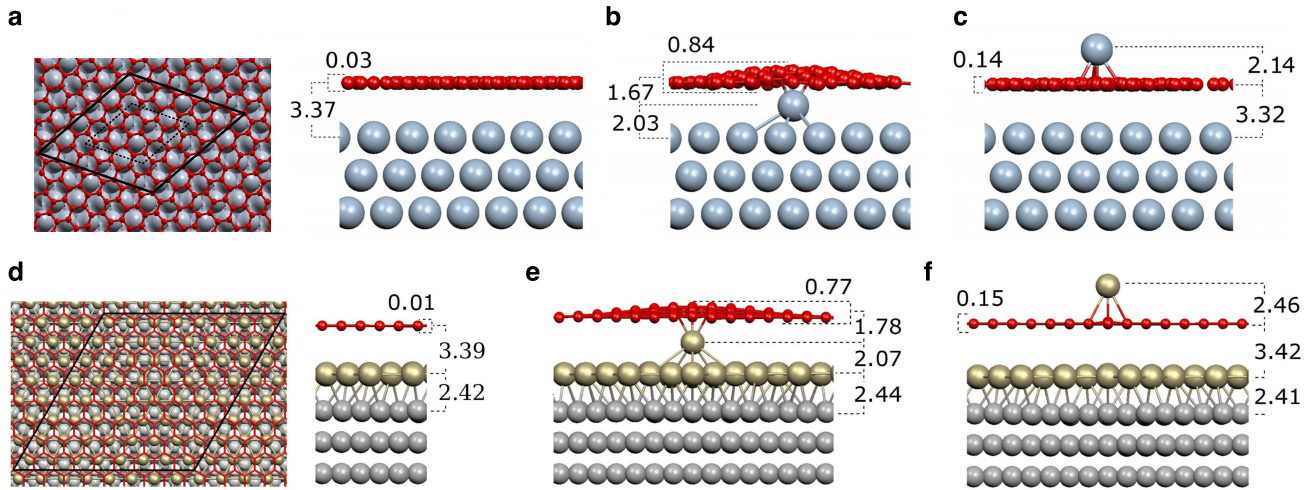}
\caption{\label{geom}
(a) Top and side view of the G/Pt(111). The ($3\times3$) supercell has been enlarged to ($6\times6$) to avoid interactions between adatoms in configurations (b) and (c). (b) Side view of G/Pt(111) with Pt adatom intercalated between graphene and the first Pt layer. (c) Same as (b), but with Pt$_{\mathrm{ad}}$ placed above (on top) of a C atom in graphene. (d) Relaxed geometry of the G/Au/Ni(111) structure. (e) Same as (b) for G/Au/Ni(111). (f) Same as (c) for G/Au/Ni(111). Graphene is represented either by red balls or sticks, while Pt, Au and Ni atoms by blue, yellow and grey balls, respectively. The black parallelograms in (a) and (d) mark the G($3\times3$)/Pt($\sqrt{7}\times\sqrt{7})R19.1^{\circ}$ and G($9\times9$)/Au($8\times8$)/Ni($9\times9$) supercells, respectively. All distances are given in angstroms.}
\end{figure*}

We have employed realistically large supercells to properly account for the the moir\'e patterns
and reconstructions known for these systems as well as to minimize the direct interaction
between the adatoms (Fig.\ref{geom}). In the case of the G/Pt(111) we
considered a thick Pt(111) slab (6 layers) with graphene adsorbed on top assuming a
G-($3\times3$)/Pt-\rethree\ supercell which corresponds to a minimal lattice mismatch.\cite{martingago}
In order to reduce the interaction between defects among neighbouring supercells we have enlarged
the $(3\times3)$ supercell to a $(6\times6)$ and placed a Pt adatom either on top of the G
in an $atop$ configuration (\ptad), or between the G and the Pt surface at an $fcc$ site and
below a C atom (\ptin). On the other hand, we modeled the G/Au/Ni(111) system assuming a
($9\times9$)/($8\times8$)/($9\times9$) commensurability between the G, Au and Ni lattices, respectively, with the Au layer intercalated between the G and the four Ni layers thick slab.
The Au adatoms have been incorporated either on top of the graphene at an $atop$ site (\auad),
or in between the G and the Au layer below a C atom and at an $hcp$ site (\auin). The final adsorption structures have been
obtained after relaxing the graphene, the adatom, and the first two metal layers
until forces were smaller than 0.04 eV/\AA. In all calculations including SOC for the
G/Au/Ni(111) systems the spin quantization axis was set along the $z$ direction (out-of-plane).
Although estimates of the magnetic anisotropy employing the force theorem indicate that the
in-plane magnetization is more favourable, we have chosen the out-of-plane orientation in order
to preserve the $p3m$ symmetry and thus facilitate the interpretation of the spin textures. The
effect of choosing a different spin quantization axis will be briefly discussed in section IV.

Finally, the electronic structures have been evaluated in the form of projected density of states PDOS\ke\, calculated for the semi-infinite surfaces constructed after replacing the bottom layers of
the slabs by a semi-inifinte bulk following the Green's functions based prescription detailed in
Refs.~\onlinecite{ysi2,loit}. Unfortunately, unfolding the G-projected band structure into its
primitive BZ is not possible in the adatom configurations since the strong interaction induce
large distortions which break the translation symmetry within the G layer. Hence, all projections
are presented folded into the supercell's BZ.

\section{G/Pt(111): intercalation vs decoration with Pt adatoms}
\begin{figure*}
\includegraphics[width=\textwidth]{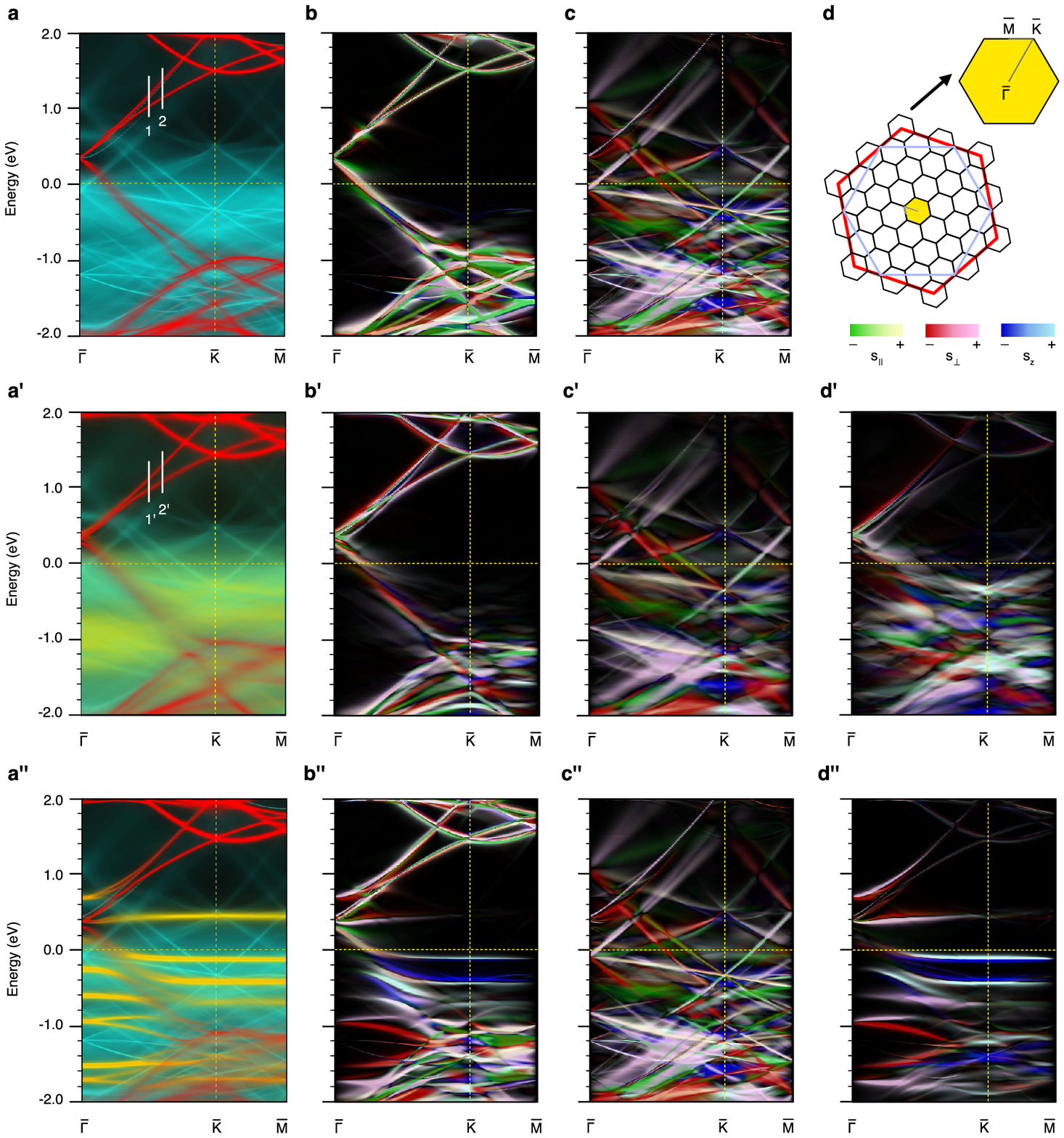}
\caption{\label{pt}
Electronic and spin structure of G/Pt(111) intercalated/decorated with single Pt atoms. (a) Density of states of G/Pt(111) calculated in ($6\times6$) supercell and projected on graphene (red) and Pt (light-blue).  (b) Corresponding spin texture projected on graphene. The color scheme is defined as follows: green/red shades refer to the direction of spin parallel/perpendicular to the momentum, while blue corresponds to the out-of-plane component; light/dark tones denotes positive/negative values of each component, see also the inset summarizing the legends in the bottom of panel (d). (c) Same as (b) projected on Pt substrate. (d) Brillouin zones of the ($6\times6$) supercell (small black hexagons), G-($1\times1$) primitive cell (red hexagon), and Pt-($1\times1$) primitive cell (blue hexagon). The selected $k$-lines are marked within the yellow hexagon in the center. (a'-c') Same as (a-c) for the configuration with intercalated Pt atom; its PDOS in (a') is colored in yellow, and its spin texture is displayed in (d'). (a''-d'') Same as (a'-d') for configuration with single Pt atoms adsorbed on top of G.
}
\end{figure*}

Figure~\ref{geom}(a-c) shows the relaxed geometries of all considered G/Pt(111) structures, that is;
the defect-free case in (a), the intercalated adatom
between G and the Pt(111) surface in (b) and the atop adatom adsorption in (c).
 Figure~\ref{pt} presents all the corresponding
electronic structures and spin textures along the high-symmetry lines of the shrinked
$(6\times6)$ BZ.

Let us first briefly summarize the main results obtained for the defect-free configuration as
a detailed study for this case has already been presented in Ref.~\onlinecite{soc_prb}.
Given the weak interaction indicated by the large G-metal distance of 3.37~\AA\ (physisorption
regime\cite{fingerprints}) the DCs can still be clearly resolved in the PDOS map in Fig.~\ref{pt}(a), where the G
(red) and surface Pt (light blue) projections have been superimposed --recall that the G's $K$ and
$K'$ points are backfolded into the supercell's $\Gamma$ point.
In Figs.~\ref{pt}(b)-(c) we present the spin textures projected on the G and the Pt surface,
respectively, where we have simultaneously plotted the three spatial components of the spin
polarization employing a different color scheme for each of them: \mpar\ green, \mper\ red and
$s_{z}$ blue tones, where \mpar\ and \mper\ correspond to the in-plane spin components
projected along the $k$-line and perpendicular to it, respectively, and $s_z$ to the out-of-plane
component. Contrary to the PDOS case, the G-Pt interaction can be clearly appreciated
in these maps via the rich spin texture induced in the DCs by the hybridization with the $d$-bands,
involving multiple spin reorientations in all the occupied states region and up to around 0.8~eV
above the Fermi level (E$_f$). Furthermore, and as shown in Ref.~\onlinecite{soc_prb}, the splitting
of the G bands is by no means uniform, attaining {\it giant} values larger than 100~meV at mini-gaps, but only a few tens of meV at most in the quasi-linear regions.

\subsection{Intercalation between graphene and Pt surface}
Intercalation of the Pt adatom (Pt$_{\mathrm{in}}$) between the G and the substrate induces a
strong buckling in the former with a corrugation as large as 0.8~\AA, with short bond lengths of
2.1~\AA\ between Pt$_{\mathrm{in}}$ and the closest carbon atoms. At the same time, the G layer is displaced
upwards so that the lowest C atoms lie 3.7~\AA\ above the Pt surface.
In such geometry, we expect a weakening of the overall interaction of the G with the Pt surface
at the expense of a stronger one with the intercalated defect.
In the PDOS map presented in
Fig.~\ref{pt}(a'), consisting of superimposed bands of G (red), Pt adatom (yellow) and the Pt
surface (light blue), the adatom contribution appears as a rather faint smudge (yellowish tones)
indicating, as expected, a strong hybridization with the continuum of Pt bulk states.
Close proximity of the C atoms with the Pt$_{\mathrm{in}}$ leads to important changes in the
DCs with respect to the defect-free case; one of the cones vanishes almost entirely below E$_f$
while the other remains well-preserved but strongly broadened in the whole considered region.

The G's spin structure, shown in panel (b'), also reveals strong differences with respect to the
defect-free case. As can be inferred from the substrate's and adatom's spin textures shown in
(c') and (d'), it now follows more closely the spin of the latter. In fact, due
to the strong G-Pt$_{\mathrm{in}}$ interaction, the Dirac point can be clearly resolved in
panel (d') as well as the strong hybridization with one of the DCs.
Surprisingly, the Pt$_{\mathrm{in}}$ spin texture is markedly different from that at
the Pt surface, which closely resembles the defect-free case (panel (c)), implying that the
SOC at the surface is hardly affected by the presence of the adatom.

\begin{figure}
    \includegraphics[width=\columnwidth]{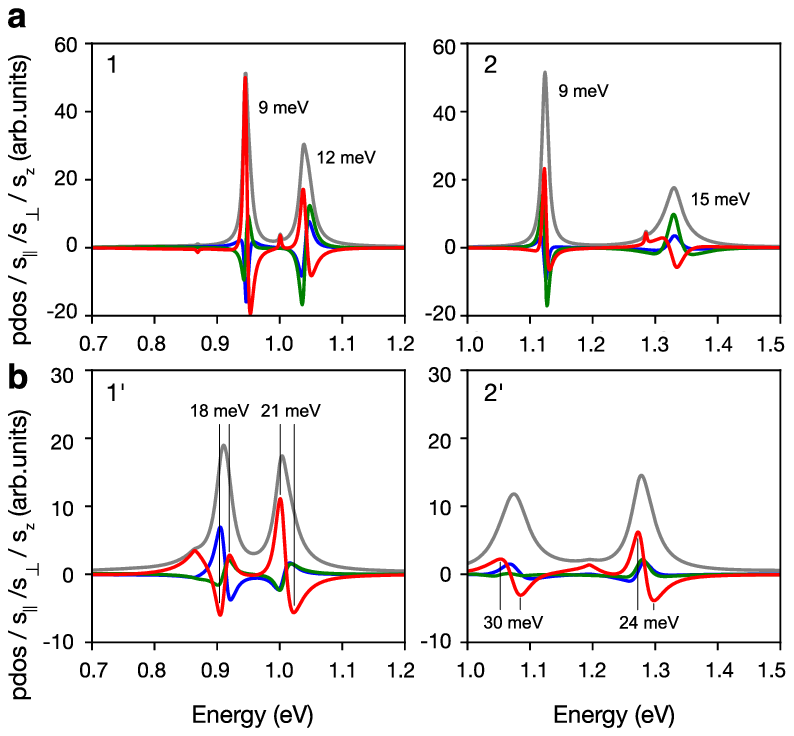}
    \caption{\label{peaks_pt}
    PDOS(E) and $\vec{s}$(E) single spectra extracted from the maps in Fig.\ref{pt} (a)-(a') at two different k-points (left-hand and right-hand panels) marked with white lines in Fig.\ref{pt}. Panel (a) corresponds to the defect free case and (b) to the intercalated model. Only unoccupied DC branches are shown. The numbers shown in the plots refer to the values of spin-orbit derived spin-splitting of the bands corresponding to each peak in PDOS(E). Grey, red, green and blue lines represent the PDOS, and \mper, \mpar\, and $s_z$ components, respectively.
    }
    \end{figure}

On the other hand, at energies above $\sim 1$~eV, there are hardly any Pt$_{\mathrm{in}}$ states
and the DCs appear at first sight very similar as in the defect free case, allowing a direct
comparison between their respective SOC induced splittings.
Figure ~\ref{peaks_pt}(b) presents single spectra corresponding to spin vector versus energy curves
$\vec{s}$(E)
extracted from panel (b') for two selected $k$-points in the empty states region (indicated by
white line segments). They are compared versus analogous data calculated for the defect-free model.
The spin-splittings are clearly larger by at least a factor of two in the case of the intercalated
model, although the PDOS (gray lines) is significantly broadened as a result of the strong
G-Pt$_{\mathrm{in}}$ interaction. Thus, Pt$_{\mathrm{in}}$
intercalation appears as a quite efficient way to enhance the spin-orbit proximity effect.

\subsection{Pt adsorption on top of G/Pt(111)}
Contrary to the intercalation case, the adsorption of a Pt adatom on top of G/Pt(111) leads to
hardly any buckling of the G with a corrugation below 0.1~\AA\ (see Fig. \ref{geom}).
However, a very short distance between the adatom and the G (2.14~\AA) induces a strong
interaction and important changes in the G's electronic structure, as can be noticed
in Fig.~\ref{pt} (a'') where the PDOS\ke\, of G, Pt$_{\mathrm{ad}}$ and Pt(111) are superimposed
following the same color scheme as in (a'). The most striking feature is the bunch of intense
localized bands belonging to the adatom (yellow) which completely tear the lower DCs and notably
alter the upper ones.
Such picture is consistent with a simpler model where the Pt surface has
been removed. Indeed, the PDOS of a pure G+Pt$_{\mathrm{ad}}$ configuration, shown in
Fig.~\ref{s1}(a) in the Appendix A, strongly resembles the one in panel (a''), indicating
that the G-Pt$_{\mathrm{ad}}$ interaction overrules that with the Pt substrate as expected from
their close proximity.
An orbital analysis of the adatom's states reveals that below E$_f$ all of them are mainly
of 5$d$ character, while only the band at approximately +400 meV, which crosses the Dirac point,
has an $sp$ origin.

The same applies to the spin textures shown in Figs.~\ref{pt}(b''-d''). The G and \ptad\
projections (panels (b'') and (d''), respectively) are highly reminiscent of their substrate-free
counterparts in Fig.~\ref{s1}(b) and (c). The quasi-atomic states at energies around $-0.1$, $-0.4$
and $-0.6$~eV can be clearly seen in the G-projected $(\vec k,E)$ map throughout the entire BZ due to their
strong hybridization; they present a spin splitting of $\sim 200$~meV and as they tear the DCs,
the $\pi$-bands are endowed with similar splittings. At each anti-crossing region their
magnetization aligns with that of the \ptad\ state (of intrinsic character) and mantain
this orientation (mainly out-of-plane, $\pm s_z$) until the next anti-crossing. The Pt surface,
nevertheless, still influences the G's spin texture, specially at energies where the \ptad\ bands
are absent: below $-1$~eV and above $+1$~eV, where the in-plane $s_\perp$ and $s_\parallel$ spin
components become patent.

\subsection{Dirac point analysis}
\begin{figure}[ht!]
\includegraphics[width=0.98\columnwidth]{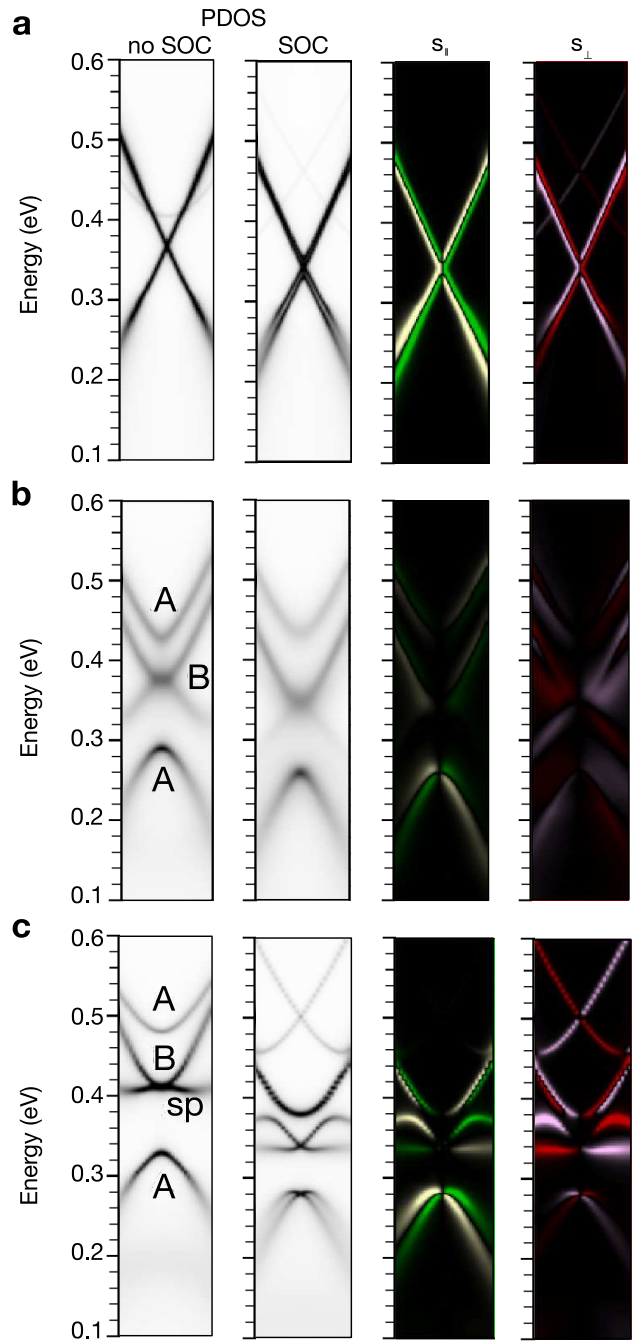}
\caption{\label{pt_dc}
G-projected electronic and spin structure around the $\Gamma$ point for the G/Pt(111) systems:
(a) defect-free, (b) intercalated
\ptin\ defect and (c) decorated \ptad\ defect. First and second columns show the PDOS
obtained without and with SOC, respectively, while right-hand columns correspond to the in-plane
spin components, $s_\parallel$ and $s_\perp$.
}
\end{figure}

 In Figure~\ref{pt_dc} we present high resolution graphene projected PDOS and $s_\parallel$ and
$s_\perp$ ($\vec k,E$) maps around the Dirac point (DP) for the three configurations considered; the $s_z$
component has been omitted since it is significanly less intense than the in-plane ones in
all cases. Additionally, and in order to visualize the role played by the SOC, in the
leftmost column we present the graphene's PDOS calculated under the scalar-relativistic
approximation. For the defect free case, panel (a), and in the absence of SOC we obtain
sharp linear $\pi$-bands and a gapless DC
consistent with the quasi-free standing character of the G. When the spin-orbit coupling
is turned on, the intrinsic SOC opens a small gap (below 10~meV) which, however, is hindered
by the broadening of the $\pi$-bands due to their hybridization with the Pt substrate. Hence, no quantum spin Hall phase is expected. On the other hand, Rashba SOC is patent in the
$s_{\parallel/\perp}$ maps with splittings of the order of 10~meV (30~meV) in the upper (lower) cones.
Furthermore, the spin texture is far from helical, having a larger $s_\parallel$ component
than $s_\perp$.

The quasi-free standing picture changes drastically for the two defected
configurations. In the intercalated case, panel (b), sublattice symmetry is broken since
the Pt adatom resides below a C atom (sublattice A), opening a large gap ($\approx 130$~meV)
between its associated DCs, while the other DP (sublattice B) remains gapless, although the
bands loose their linear behaviour. Furthermore, the G's PDOS intensity is significanly
smaller than in the defect-free case due to the reduced C-Pt distance.
The main effect of the SOC here is an
increase in the gap for DP-A and of the Rashba splitting of all cones. This is particularly clear in the
lower DC-A, where the splittings attain values close to 40~meV.
When the Pt adatom is adsorbed on top of a
C atom, panel (c), the sublattice symmetry is again broken and a gap larger than 150~meV opens at the DP-A. On the other hand, the lower DC associated to the sublattice B is destroyed
due to the presence of the \ptad\ $sp$ atomic level at around 0.4~eV. Apart from the Rashba
splitting of the lower DC-A (larger than 40~meV) and, to a less extent, of the upper DC-B, SOC
induces a splitting of the adatom's $sp$ state, so that one component remains flat
(at around 0.34~eV) while the other bends as it anti-crosses the DC.

\section{Adsorption of single Au adatoms in G/Au/Ni(111)}
\begin{figure*}
\includegraphics[width=\textwidth]{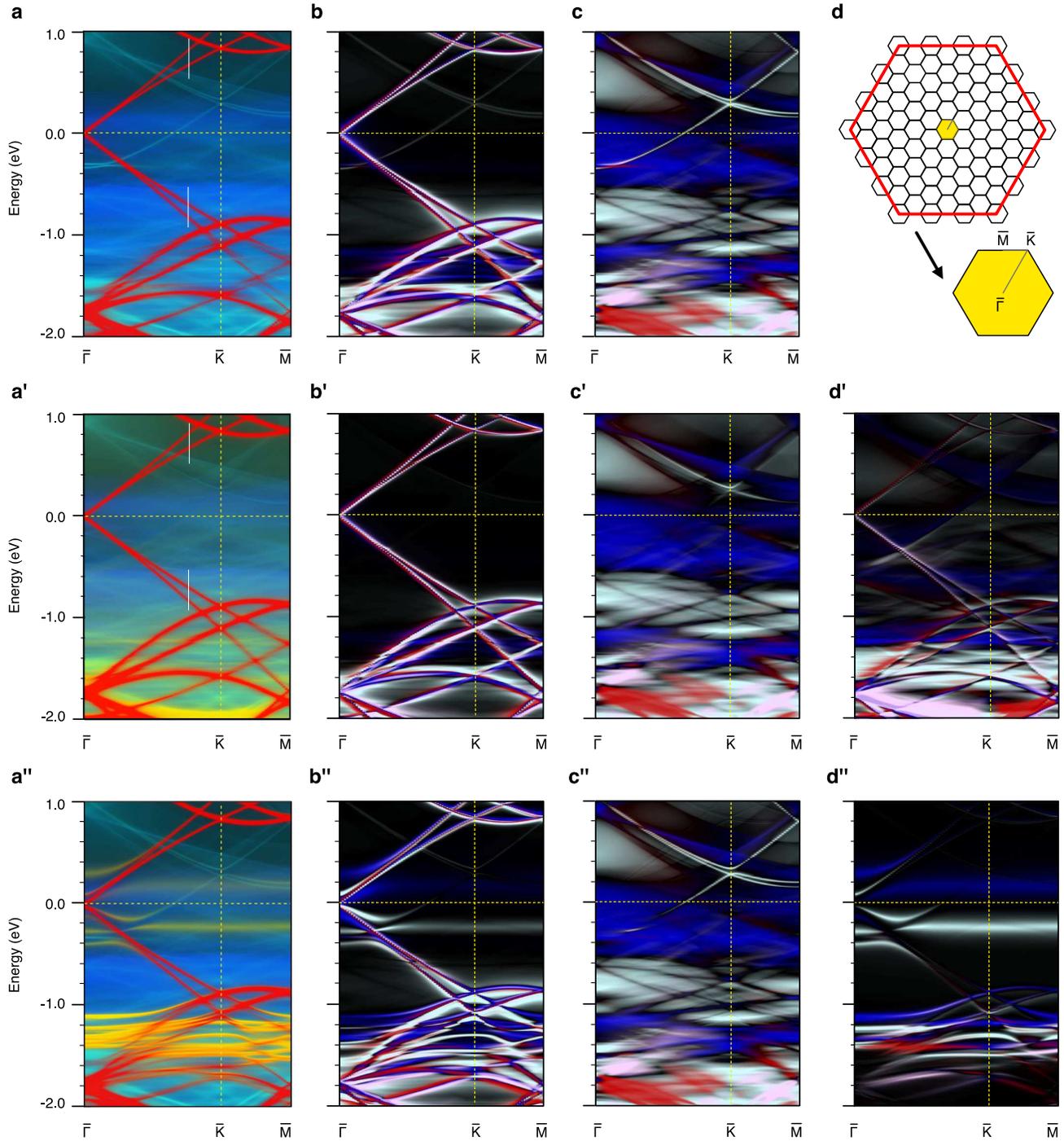}
\caption{\label{au}
Electronic and spin structure of G/Au/Ni(111) intercalated/decorated with single Au atoms. (a) Band structure of G/Au/Ni(111) along $\Gamma-K-M$ in folded $(9\times9)$ BZ represented as PDOS\ke projected on graphene (red), gold (light blue) and Ni surface (dark blue) superimposed at one map. (b) Corresponding graphene's spin texture after superimposing the $x/y/z$ components, each color coded as explained in Fig.\ref{pt}. (c) Same as (b), but projected on intercalated Au layer. (d) Brillouin zones of the  ($9\times9$) supercell (small black hexagons), and G-($1\times1$) primitive cell (red hexagon); the considered $k$-lines are labeled within the yellow hexagon. (a'-c') Same as (a-c) for the configuration with additional Au atom intercalated below the G; yellow shades in (a') denote its PDOS, while panel (d') shows its spin texture. (a''-d'') Same as (a'-d') for the configuration of G/Au/Ni(111) with Au atoms adsorbed on top of the G. The spin textures projected on Ni(111) are neglected in all cases.
}
\end{figure*}

\begin{figure}
    \includegraphics[width=\columnwidth]{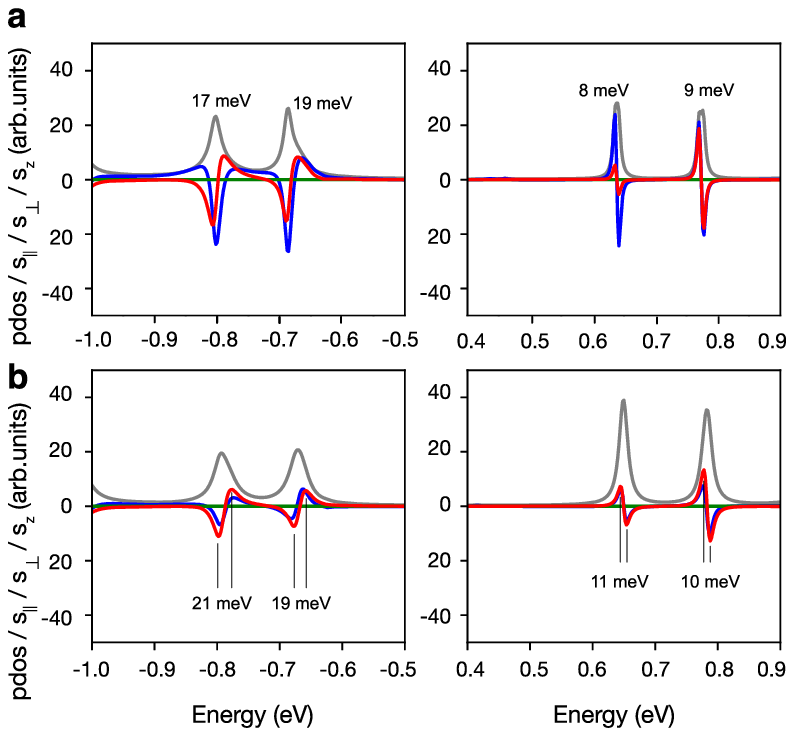}
    \caption{\label{peaks_au}
    PDOS(E) and $\vec{s}$(E) single spectra extracted from the maps in Fig.\ref{au} (a)-(a') at specific k-point marked with white lines. Panel (a) corresponds to the defect-free case and (b) to the model containing additional intercalated Au atom. Left-hand (right-hand) panel shows occupied (unoccupied) DC branches. The numbers shown in the plots refer to the values of spin-orbit derived spin-splitting of the bands corresponding to each peak in PDOS(E). Grey, red, green and blue lines represent the PDOS, and \mper, \mpar\, and $s_z$ components, respectively.
    }
    \end{figure}
The relaxed geometries for the G/Au/Ni(111) system are shown in Figures~\ref{geom}(d), (e) and (f)
for the defect-free case, the intercalated \auin\ adatom and the adatom \auad
on top of the G,
respectively. In the former, the weak G-Au interaction~\cite{soc_prb,gold-prl,klusek,gold-dft,gold-domains,gold-rpa}
leaves an uncorrugated graphene layer lying 3.4~\AA\ above the metal surface.
Figures~\ref{au}(a-c) summarize its associated electronic and spin structure along the
high-symmetry lines of the supercell's BZ.\cite{soc_prb}.
Overall, the hybridization between graphene and the underlying Au/Ni(111) is weaker than in
G/Pt(111) case.
In the combined PDOS\ke\, map (a), the G (red), Au (light blue) and Ni surface
(dark blue) projections have been superimposed. The quasi-freestanding character of the G
manifests in almost undoped and well-preserved DCs down to binding energies of around $-1$~eV,
in agreement with previous experimental works.\cite{first13meV, main_nat}
The most intense Ni related features are located at approximately $-0.6$ and $+0.1$~eV,
corresponding to the top of the majority and minority $d$-bands, respectively. Several
gold $sp$ bands (the most prominent of them the Shockley-type surface state~\cite{soc,soc_exp}
(SS) emerging from $\Gamma$ at $-0.33$~eV) cross the BZ whereas fingerprints of the Au
5$d$-bands (light blue) appear below $-1$~eV distorting the DCs.

In spite of the fact that this system is magnetic and, hence, there exists an interplay between
SOC and exchange interactions, the G's spin texture shown in Fig.~\ref{au}(b) appears far
less complex than in the G/Pt(111) case. Indeed, in the [-1,+1]~eV range where the DCs appear
almost intact, their spin vector has only two components both perpendicular to the momentum~:\cite{soc_prb}
an in-plane helical component arising solely from the SOC, $s_\perp$, and an
out-of-plane one, $s_z$, mainly induced by the Ni magnetic order.
It is also noteworthy the different broadenings of the spin-splitted branches,
particularly around the DP at $\Gamma$ where the minority (dark blue) component is much broader than
the minority one (light blue) whereas along $K-M$ and at around $-0.9$~eV the opposite behavior holds.
The $\pi$-band splittings in this energy window are only of the order of 10 meV, in
agreement with previous experimental data\cite{first13meV} and several theoretical results\cite{main_nat,soc_prb, voloshina}.
However, the helical spin texture should not hold anymore when domains with
different in-plane magnetizations are present at the Ni surface, as expected in real
samples\cite{main_nat}. In such case, one may still expect that the values of the splittings will
remain small since their magnitude is mainly related to the magnetic coupling between the G and the
Au/Ni(111) surface rather than to the SOC.

The spin projected on the intercalated Au layer (panel (c)), on the other hand, mainly reflects the
hybridization with the Ni(111) spin-polarized bands again displaying light (majority) and dark
 blue (minority) regions.
SOC manisfests most notably in the lower energy region (below $-1.0$~eV), where large in-plane
components (red) can be clearly seen at several energies.

\subsection{Intercalation with single gold atoms}

We again explored the role of adatoms either adsorbed above the graphene or intercalated between
the graphene and the Au monolayer. The relaxed structures, shown in Figs.~\ref{geom}(e) and (f),
follow analogous trends as in the G/Pt system. The intercalated adatom
induces a significant buckling in the G (0.8~\AA) while its average distance to the top Au layer
is significantly increased from 3.4~\AA\ to 3.9~\AA.
The associated PDOS and spin \ke\ maps are presented in the middle
panels in Fig.~\ref{au}.
As shown in (a') where the additional \auin\ projection is colored in yellow, and
in contrast to the G/Pt(111) case, the intercalated adatom
introduces only subtle changes in the band structure
(e.g. removal of the Au's SS) leaving the graphene's DCs hardly affected. The contribution
of the highly delocalized \auin\ $sp$ states covers most of the map as can be seen by the change
in the blue tones compared to the defect-free configuration in panel (a), while intense
$d$-states appear below $-1$~eV showing little dispersion. We also note that \auin\ shows
no significant spin-polarization (below 0.01~$\mu$B) when intercalated.
The spin textures projected on the G and the gold surface layer, panels (b') and (c'), respectively,
are very similar to their defect-free counterparts ((b) and (c)), implying that the adatom has
little impact on them. The main difference is a reduction of the $\pi$-band
broadening due to the enlarged G-Au average distance. In Fig.~\ref{peaks_au} we compare G-projected DOS$(E)$ and $\vec s(E)$ curves between the
defect-free (a) and the intercalated (b) cases for both the lower and upper DCs
at a representative $k$-point (marked by the white segments in Figs.~\ref{au}a-a').
There are only very small changes (a few meV) in the splittings between both systems, with values
of $\sim 10$~meV in the upper cones and $\sim 20$~meV in the lower ones.
Therefore, intercalation of an Au adatom hardly enhances the SOC derived spin splitting in
the G/Au/Ni(111) system, in contrast to the G/Pt(111) case. We assign this difference to the
absence of \auin-$d$ states close to E$_f$.

\subsection{Decoration with single gold atoms}

When Au$_{ad}$ is adsorbed on top of the G, the latter remains hardly
corrugated (0.15 \AA), while the C-\auad\ bond distance becomes very short (2.46 \AA).
Below $-1$~eV, the atomic-like \auad\ $d$-states (intense yellow in (a'')) strongly hybridize
with the DCs opening multiple gaps. Moreover, the most relevant
feature is the pair of flat bands that run above and below $E_f$ and which strongly perturb and
tear the $\pi$-bands close to the DP.
As can be clearly seen in the G and \auad\ spin projections of panels (b'') and
(d''), each band holds opposite spins with only $s_z$ component. An orbital
analysis reveals that they correspond to the 6$s$ state of \auad\ which is
exchange splitted by $\sim 0.4$~eV and, in analogy with an Au isolated atom,
is responsible for the adatom's spin polarization (the total \auad's magnetic
moment is 0.56~$\mu$B).
The analogous calculation for the simpler G+Au$_{\mathrm{ad}}$ model
(that is, after removing the Au/Ni(111) surface), shown in Figure~\ref{s2}, yields a
very similar band and spin structure, indicating that the spin polarization of
the \auad\ atom is unrelated to that of the substrate. This is further
corroborated by the fact that the spin texture projected on the Au layer (c'')
is almost identical to that of the defect-free case (c).


\subsection{Dirac point analysis}
\begin{figure}[ht!]
\includegraphics[width=0.99\columnwidth]{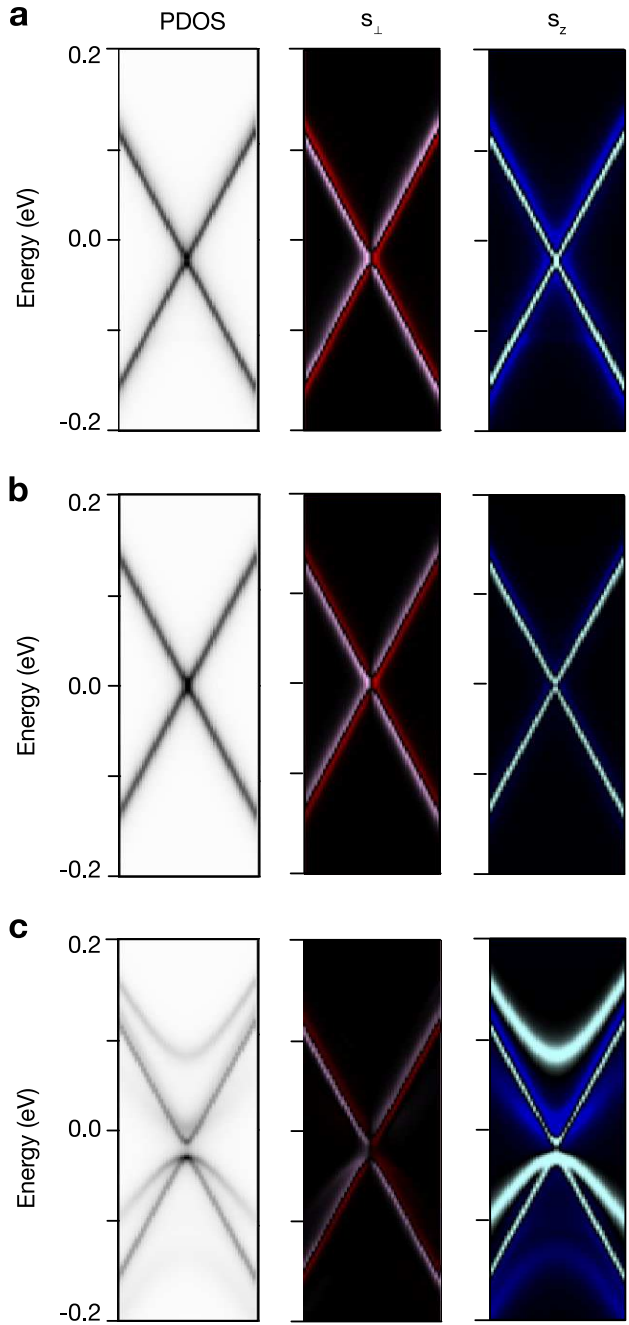}
\caption{\label{au_dc}
G-projected electronic and spin structure around the $\Gamma$ point for the G/Au/Ni(111) systems:
(a) defect-free, (b) intercalated
\auin\ defect and (c) decorated \auad\ defect. Left, center and right columns show the PDOS,
as well as the $s_\perp$ and $s_z$ spin components, respectively.
}
\end{figure}

 High-resolution graphene projected PDOS and $s_\perp$ and $s_z$ ($\vec k,E$) maps are displayed
in Figure~\ref{au_dc} for the three G/Au/Ni(111) configurations (this time the $s_\parallel$
component is negligible in all cases). The equivalent PDOS and $s_z$ maps calculated neglecting
SOC have been omitted since they are visually identical to those shown in the figure. Therefore,
as stated above, the role of SOC in this system is mainly to introduce an $s_\perp$ spin
component (helical spin texture). Siimilar to the G/Pt case, the defect-free configuration
presents quasi-perfect DCs while the possible presence of a small gap due to intrinsic SOC is
masked by the broadening of the $\pi$-bands. The broadening, in fact, is much larger for the
$-s_z$ bands (dark), as could be expected from the fact that the Au/Ni(111) PDOS
around the Fermi level is mainly occupied by the minority Ni bands.

As shown in panel (b), intercalation of \auin\ hardly alters the DP or its
spin texture due to the low Au PDOS around the $E_F$ (the adatom's $d$-states all lie
at binding energies below $-1$~eV). However, the situation is drastically different when
the G is decorated by the adatom (panel (c)). The interaction between the spin-splitted
\auad\ $s$-levels induces a large gap in the DCs associated to sublattice A (the C atom below
\auad) which are also spin-splitted (bright and dark parabolas in the $s_z$ map). In contrast,
the DCs of the sublattice B remain fairly linear except for a small gap at $\Gamma$.

\section{Summary and conclusions}
We have investigated the spin-orbit proximity effect in graphene on metallic substrates decorated
or intercalated by metallic adatoms focusing on two specific graphene/metal systems, non-magnetic
G/Pt(111) and magnetic G/Au/Ni(111) previously studied experimentally. \cite{platinum,platinum2,first13meV,main_nat,krivenkov}
Depending on the location of the adatom, two very different scenarios are reached; adsorption on top
leaves the graphene essentially uncorrugated but hybridizations with the atomic-like $d$-states
leads to densely teared $\pi$ bands resembling freestanding graphene decorated by adatoms. It turns
out that in the two systems considered the \ptad\ and \auad\ adatoms present states close to
E$_f$, thus the quasi-linear parts of the DCs close to the DP are largely distorted and the
electronic structure of G loses its linear character.

On the other hand, when intercalated between the graphene and the metal surface the former
becomes highly corrugated making short bonds with the adatom but with an average distance to the
surface larger (by $\sim 0.4$~\AA) than in the defect-free case. In this geometry, the
adatom's states strongly hybridize with the substrate's continuum of bands largely losing
their atomic-like character and therefore, their effect on the $\pi$-bands is less intense than
for adsorption on top. In G/Pt(111) the upper DCs remain almost unaltered exhibiting
a SOC-induced complex spin texture similar to the defect-free case. Interestingly, the close
proximity of the G to the \ptin\ leads to an
increase in the $\pi$-band splittings in the empty states region by up to a factor of three.
This is not the case, however, for G/Au/Ni(111) which presents similar splittings as in the
defect-free case since the \auin\ $d$-states lie at higher
binding energies and their impact on the upper DC is less significant.

A detailed analysis of the G's Dirac point shows
that the role of intrinsic SOC is minimal in all configurations, inducing gaps smaller than the
broadening of the $\pi$-bands; this is an expected result since in all the defected configurations
the G's sublattice symmetry is broken.\cite{weeks} Therefore, the proximity effect
in the systems under consideration relies mainly on the Rashba-type SOC transfer.

Finally, we recall that the G/Au/Ni(111) system is the most puzzling one, since two
very different spin-splittings for the $\pi$-bands have been reported: around 10~meV~\cite{first13meV} and
giant values close to 100~meV~\cite{main_nat}. A subsequent STM study~\cite{krivenkov}, including simplified
theoretical models, tentatively assigned the small splittings to a full gold monolayer, while the
giant values would correspond to sub-monolayer phases where small Au clusters or even individual atoms
lie intercalated between the Ni(111) surface and the G. Furthermore, the Ni surfacemost layer was shown
to be reconstructed presenting a misfit dislocation loop structure.~\cite{interface} All our models considered,
based on a full gold monolayer (plus an adatom), and even taking into account the reconstructed
Ni(111) surface~\cite{soc_prb}, always yield small splittings of the order of 10~meV which are
driven by the substrate's spin polarization.
Therefore, even if the giant splittings come from
gold sub-monolayer phases or a different type of Ni-Au surface alloying (which we have not considered), we
believe their magnitudes are determined by the magnetic coupling with the metal surface and not by the SOC.

\begin{acknowledgments}
J.S. acknowledges Polish Ministry of Science and Higher Education for the Mobility Plus Fellowship (Grant No. 910/MOB/2012/0). J.I.C acknowledges support from the Spanish Ministry of Economy and
Competitiveness under contract Nos. MAT2015-66888-C3-1R and RTI2018-097895-C41. Part of the
calculations have been done using the supercomputer facilities at the Barcelona Supercomputing Center under the activity
ID QCM-2015-2-0008.
\end{acknowledgments}

\appendix
\section{Graphene decorated by single metal adatoms}
\begin{figure}[ht]
\includegraphics[width=0.5\textwidth]{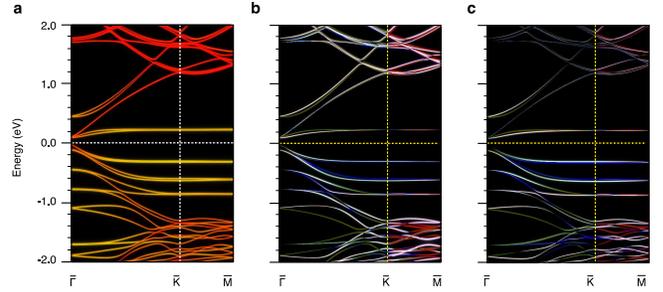}
\caption{\label{s1} (a) DOS\ke\ map projected on the G (red) and Pt
adatom (yellow). No substrate was included in this case. (b) Spin texture corresponding to graphene's PDOS presented in (a). (c) Same as (b) projected on Pt adatom. Color scheme same as in Figs \ref{pt} and \ref{au}.
}
\end{figure}

\begin{figure}[ht]
\includegraphics[width=0.5\textwidth]{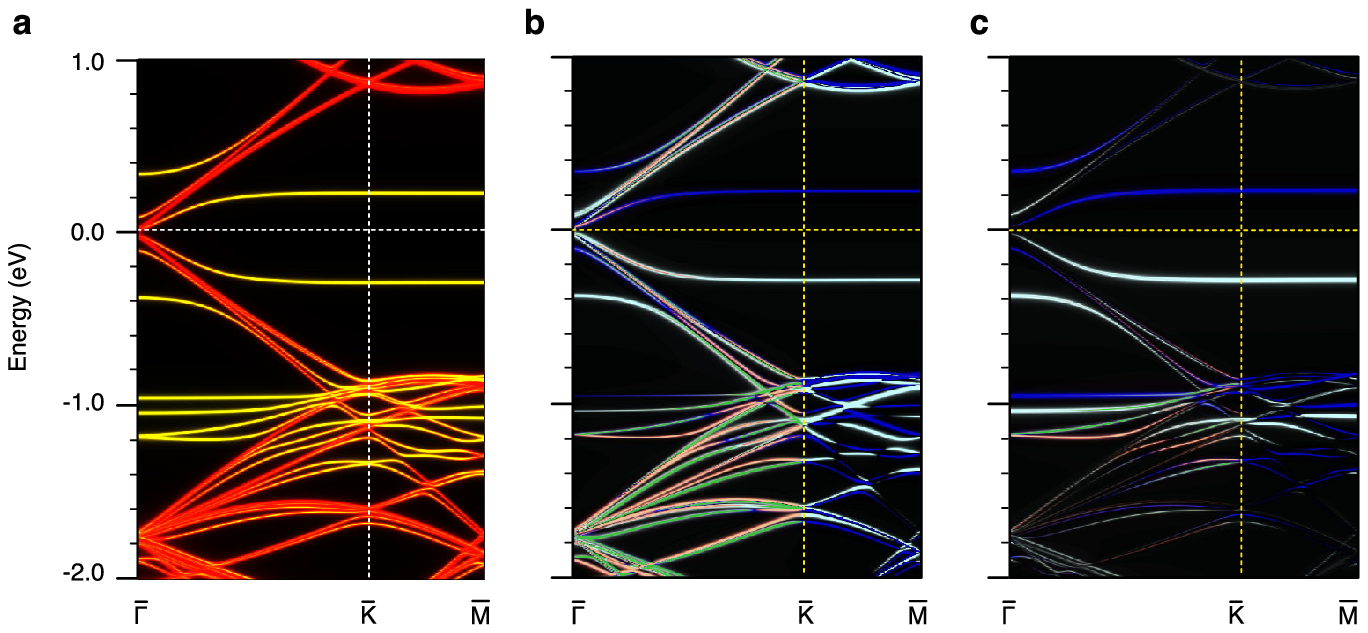}
\caption{\label{s2} (a) DOS\ke\ map projected on the G (red) and Au
adatom (yellow). No substrate was included in this case. (b) Spin texture corresponding to graphene's PDOS presented in (a). (c) Same as (b) projected on Au adatom. Color scheme same as in Figs \ref{pt} and \ref{au}.
}
\end{figure}

Although pure decoration of graphene with single metal adatoms was widely studied in the literature and is well understood in terms of model Hamiltonians,\cite{fabian-tightbinding, fabian_copper, mertig} we present below DFT calculations without the substrates employing the same supercells as considered for the systems discussed in the main text. Figures \ref{s1} and \ref{s2} show the electronic and spin properties of graphene decorated by a
single Pt and Au atom, respectively, without including any metallic surface.\cite{calleja,main_nat,carbon,mertig} In both cases, the overall picture is similar to the analogous configuration on top of the metallic substrate, which
confirms that the graphene-adatom interaction becomes dominant.
We can easily observe in Fig.~\ref{s1} that the states of the adatom strongly interact with the DCs opening a $\sim 100$~meV band gap and several anticrossing gaps below $E_F$. Comparing with
Fig.\ref{pt} (a''), we can conclude that the only effect of the metallic substrate is the $p-$type
doping of $\sim 300$~meV and the broadening of the bands due to the interaction with several
substrate's states. In the case of G/Au/Ni system (Fig. \ref{s2}) the interaction between graphene and Au adatom induces similar changes in the DCs, but given the smaller number of Au states close to the Fermi level, the DCs are less perturbed than in case of the decoration with Pt atom. From comparison with Fig. \ref{au} (a'') it is clear that the substrate plays hardly any role; this behavior is quite expected as graphene can be considered quasi-freestanding on Au/Ni.


\end{document}